\pdfoutput=1

\documentclass[11pt,twoside,a4paper,cmspaper,final,collab]{cms-tdr}
\def\svnVersion{163620:163853}\def\cmsCernNo{2012-332}\def\cmsCernDate{\today}\def\cmsMessage{Submitted to Physical Review Letters}

\begin{document}\cmsNoteHeader{SUS-11-024}

\hyphenation{had-ron-i-za-tion}
\hyphenation{cal-or-i-me-ter}
\hyphenation{de-vices}

\RCS$Revision: 163620 $
\RCS$HeadURL: svn+ssh://alverson@svn.cern.ch/reps/tdr2/papers/SUS-11-024/trunk/SUS-11-024.tex $
\RCS$Id: SUS-11-024.tex 163620 2013-01-07 19:00:34Z smaria $
\cmsNoteHeader{SUS-11-024} 
\newlength\cmsFigWidth
\ifthenelse{\boolean{cms@external}}{\setlength\cmsFigWidth{0.85\columnwidth}}{\setlength\cmsFigWidth{0.4\textwidth}}
\ifthenelse{\boolean{cms@external}}{\providecommand{\cmsLeft}{top}}{\providecommand{\cmsLeft}{left}}
\ifthenelse{\boolean{cms@external}}{\providecommand{\cmsRight}{bottom}}{\providecommand{\cmsRight}{right}}
\title{Inclusive search for supersymmetry using  the razor variables in pp collisions at \texorpdfstring{$\sqrt{s} = $7\TeV}{sqrt(s) = 7 TeV}}

\date{\today}

\abstract{An inclusive search is presented for new heavy particle
  pairs produced in $\sqrt{s} = 7\TeV$ proton-proton collisions at the
  LHC using $4.7\pm0.1\fbinv$ of integrated luminosity.
The selected events are analyzed in the 2D \textit{razor} space of $M_R$,
  an event-by-event indicator of the heavy particle mass scale, and
  $R$, a dimensionless variable related to the missing transverse energy.
  The third-generation sector is probed using the event
  heavy-flavor content.  The search is sensitive to generic
  supersymmetry models
with minimal assumptions about the
  superpartner decay chains. No excess is observed in
  the number of events beyond that predicted by the standard model.
  Exclusion limits are derived in the CMSSM framework as well as for
  simplified models.  Within the CMSSM parameter space considered,
  gluino masses up to 800\GeV and squark masses up to 1.35\TeV are excluded at 95\% confidence level
  depending on the model parameters. The direct production of pairs of stop or sbottom
  quarks is excluded for masses as high as 400\GeV.}

\hypersetup{%
pdfauthor={CMS Collaboration},%
pdftitle={Inclusive search for supersymmetry using  the razor variables in pp collisions at sqrt(s) = 7 TeV},%
pdfsubject={CMS},%
pdfkeywords={CMS, LHC, SUSY, Razor, top+X, W+jets, Z+jets, QCD scaling}}

\maketitle 

Models with softly broken supersymmetry
(SUSY)~\cite{Ramond,Golfand,Volkov,Wess,Fayet} predict heavy
superpartners of the standard model (SM) particles. Experimental
searches for $R$-parity~\cite{bib-rparity} conserving SUSY at the
Tevatron~\cite{:2007ww, Aaltonen:2008rv} and the Large Hadron Collider
(LHC)~\cite{:2012mfa, Chatrchyan:2012rg, Collaboration:2011xk, daCosta:2011qk, Aad:2011hh,
Aad:2011xm, RA2, alphaT, :2011wb, Chatrchyan:2011bz,
Chatrchyan:2012qka, Chatrchyan:2012mea,Chatrchyan:2012sa,:2012th,:2012jx,
ATLAS:2012ai,ATLAS:2012ah,ATLAS:2012ag,Aad:2012rt,:2012cwa,Aad:2012cz,bib-atlas-razor} have focused on
signatures combining energetic hadronic jets and leptons from the
decays of pair-produced squarks and gluinos, with large missing
transverse energy (\ETm) from the two weakly interacting lightest,
neutral superpartners (LSPs) produced in separate decay chains.

In SUSY models, the scale of soft SUSY breaking is related to the scale of
electroweak symmetry breaking. This implies either that the soft-breaking
mass parameters cannot be too large, or that the smallness of the electroweak
scale is explained by large cancellations arising from relations among these parameters
in the high energy theory. The latter possibility is complicated
by large radiative corrections, particularly those induced by the soft-breaking parameters that are
responsible for the masses of the stops and sbottoms, the superpartners of the
third-generation quarks. It is thus of special importance to search
for the lightest allowed stops and sbottoms, whose decays will be enriched in
heavy flavor.

In this Letter we present results of an
inclusive search for new heavy particles. The analysis is
designed to be largely independent of the details of
the decay chains
and  measures
deviations from the characteristic distributions of the
relevant SM processes in the razor variable plane~\cite{rogan,razor2010}.
It is generically sensitive to
the production of pairs of heavy particles, provided that the decays
of these particles produce significant {\ETm}, that these particles
are substantially heavier than any SM particle, and that they are
strongly produced in high-energy proton-proton collisions.  The
selection requires only two or more energetic reconstructed
calorimeter objects~\cite{razor-2012PRD}. The selected events are
sorted hierarchically into exclusive data samples, categorized
according to the lepton multiplicity in the event. The analysis is
repeated with the requirement of the presence of a bottom-quark jet (\cPqb-jet) to  search for
third-generation-enhanced SUSY signatures. While in  Monte
Carlo studies all SM processes are considered, the major backgrounds
are top production and vector boson production in association with
jets.

The razor kinematic variables are based on the generic process of the pair
production of two heavy particles, each decaying to an undetected
particle plus visible decay products.  Regardless of its complexity,
each event is treated as a dijet-like event by grouping all hadronic
jet candidates and isolated electrons in the event into two
megajets~\cite{razor2010}. The razor kinematic variables are used to
test, event-by-event, the hypothesis that the megajet topology
represents the visible portion of the decays of two heavy particles.
Assuming the pair of megajets accurately reconstructs the
visible portion of the parent particle decays, the signal kinematics
is equivalent, for example, to pair production of heavy squarks
$\PSQ_1$, $\PSQ_2$, with $\PSQ_i\rightarrow j_i\PSGcz $, where the $\PSGcz$ are LSPs and
$j_i$ denotes the visible products of the decays.

The $M_\mathrm{R}$  razor kinematic variable
is defined in terms of the momentum of
the two megajets as
\begin{math}
 M_\mathrm{R} \equiv
 [
(|\vec{p}^{~j_{1}}|+|\vec{p}^{~j_{2}}|)^2 -
(p^{~j_1}_z+p^{~j_2}_z)^2]^{1/2}
\end{math}
and is, by construction, invariant under longitudinal boosts. In the
 approximation of massless megajets and negligible initial-state
 $\pt$, $M_\mathrm{R}$ equals $\gamma_\Delta M_\Delta$, where $M_\Delta \equiv
 (M_{\PSQ}^{2}-M_{\PSGc}^{2})/M_{\PSQ}$ is twice the
 magnitude of the momentum of either megajet in the respective squark
 rest frame, and $\gamma_\Delta$ is the boost factor from the
 center-of-mass frame to the squark rest frames. Note that this
 definition of $M_\mathrm{R}$ is amended from that in~\cite{razor2010} to
 avoid configurations where the razor variable is ill-defined due to
 unphysical Lorentz transformations.

The razor
observable  $M_\cmsSymbolFace{T}^\mathrm{R}$ is defined as
\begin{math}
M_\cmsSymbolFace{T}^{R}\equiv \left[ \tfrac{1}{2} \left(\ETmiss(\pt^{j1}+\pt^{j2}) -
    \vec{E}_\cmsSymbolFace{T}^{\text{miss}} {\cdot}
    (\ptvec^{\,j1}+\ptvec^{\,j2})\right) \right]^{1/2}\!,
\end{math}
where $\ptvec^{\,j1,2}$ are the transverse momentum vectors of the
two megajets and $\vec{E}_\cmsSymbolFace{T}^{\text{miss}}$ is the missing
transverse  momentum vector (also referred to as missing tranverse energy).
The razor dimensionless ratio is defined as
\begin{math}
R \equiv \frac{M_T^R}{M_\mathrm{R}}
\;.
\end{math}
For signal events $M_\cmsSymbolFace{T}^\mathrm{R}$ has a maximum value (a kinematic endpoint)
of
$M_\Delta$, so $R$ has a maximum value of approximately one. Thus
signal events are characterized by a distribution in $M_\mathrm{R}$ that peaks
around $M_\Delta$, and a distribution in $R$ that peaks around 0.5, in
stark contrast with, for example, QCD multijet background events,
whose distribution in either $R$ or $M_\mathrm{R}$ is exponentially suppressed
away from zero~\cite{razor2010, razor-2012PRD}.  These properties determine a region
of the 2D razor space where the standard model background is reduced while the signal is retained.

A detailed description of the CMS detector can be found
elsewhere~\cite{CMS:2008zzk}.
A superconducting solenoid provides an axial magnetic field of 3.8\unit{T}.
 The silicon pixel and strip
tracker, the high-resolution crystal electromagnetic calorimeter (ECAL), and the
brass/scintillator hadron calorimeter (HCAL) are contained within the
solenoid. Muons are detected in gas-ionization chambers embedded in
the steel return yoke.
The HCAL, combined with the ECAL, measures
the jet energy with a resolution $\Delta E/E \approx 100\,\% /
\sqrt{E/\GeV} \oplus 5\,\%$. CMS uses a coordinate system with the origin
located at the nominal collision point, and the
pseudorapidity is defined as $\eta = -\ln [\tan(\theta / 2)]$, where
the polar angle $\theta$ is defined with respect to the
counterclockwise beam direction.

The analysis uses a set of dedicated triggers that apply lower
thresholds on the values of $R$ and $M_\mathrm{R}$ computed online from the
reconstructed jets and {\ETm}. Three trigger categories are
used: i) hadronic razor triggers applying threshold requirements
\cite{razor-2012PRD}  on
$R$ and $M_\mathrm{R}$ in events with at least two jets of $\pt>56\GeV$; ii) muon
razor triggers that have looser $R$ and $M_\mathrm{R}$ requirements than the
hadronic triggers and combined with at least one muon in the central
part of the detector (barrel) with $\pt>10\GeV$; iii) electron razor triggers with
similar $R$ and $M_\mathrm{R}$ requirements to those
  used for muons and with at least one electron of
$\pt>$ 12\GeV satisfying loose isolation criteria.
In addition, a set of non-razor
triggers is used to
define control data samples. 

Events, after detector- and beam-related
noise cleaning, are required to have at least one high-quality reconstructed
interaction vertex \cite{TRK-10-005}. When multiple vertices are
found, the one with the highest associated $\sum_\text{track}\pt^2$ is
selected.
The electron and muon candidate reconstruction and identification
criteria are described in Ref.~\cite{EWK-PAS}.  Electrons and
muons are required to lie within $\vert\eta|< 2.5$ and 2.1,
respectively, and to satisfy the identification and selection
requirements from \cite{EWK-PAS}.
Jets are reconstructed from calorimeter energy deposits
using the infrared-safe anti-\kt algorithm~\cite{antikt} with
radius parameter 0.5. Jets are corrected for non-uniformities of
the calorimeter response using energy- and $\eta$-dependent correction
factors. Only jet candidates with $\pt> 40\GeV$ within $|\eta| <
3.0$ are retained.  The jet energy scale uncertainty for these corrected
jets is 5\%~\cite{JES}. To match the trigger requirements, the $\pt$ of
the two leading jets is required to be greater than 60\GeV.
The  transverse momentum imbalance in the event, $\vec{E}_\cmsSymbolFace{T}^{\text{miss}}$,  is
reconstructed using the particle flow algorithm~\cite{PFMET}.

The reconstructed
jets
are grouped into two megajets \cite{razor-2012PRD}.
The megajets are constructed as a sum of the
four-momenta of their constituent objects.
After the baseline selection and calculation of the variables $R$ and
$M_\mathrm{R}$, the events are assigned to one of six final state
boxes according to whether the event has zero, one, or two isolated
leptons, divided according to lepton flavor (electrons and muons) as
shown in Table~\ref{tab:boxes}.

\begin{table}[ht!]
\topcaption{Razor boxes definition. The variables and requirements are
explained in the text.  \label{tab:boxes}}
\centering
\begin{scotch}{l c}
  \multicolumn{2}{c}{Lepton boxes $M_\mathrm{R} > 300$
GeV, $0.11< R^2 < 0.5$ }\\\hline
ELE-MU   & $\pt^{\Pe} > 20$\GeV, $\pt^{\mu} > 15$\GeV\\
MU-MU  & $\pt^{\mu 1} > 15$\GeV, $\pt^{\mu 2} > 10$\GeV\\
ELE-ELE  & $\pt^{\Pe1} > 20$\GeV, $\pt^{\Pe2} > 10$\GeV\\
MU  & $\pt^{\mu} > 12$\GeV\\
ELE  & $\pt^{\Pe} > 20$\GeV \\ \hline
  \multicolumn{2}{c}{HAD box $M_\mathrm{R} > 400$\GeV, $0.18< R^2 < 0.5$ }\\
\end{scotch}
\end{table}

The requirements given in Table~\ref{tab:boxes} define the full
analysis regions of the $R^2$-$M_\mathrm{R}$ plane, where the analysis is
performed for each box.  They are the loosest possible requirements
that allow for the valid background description, while at the same
time maintaining fully efficient triggers.
To prevent ambiguities for events satisfying the selection
requirements of more than one box~\cite{razor-2012PRD}, the boxes are arranged in a
predefined hierarchy, as given in Table~\ref{tab:boxes}.
Each event is uniquely assigned to the first
box whose criteria are satisfied by the event.

Six additional boxes are formed for events with at least one \cPqb-jet
tagged using the Track-Counting High-Efficiency (TCHE) \cPqb-tagging
algorithm with 1\% misidentification rate~\cite{btag,btv-11-004}.
These six boxes define the razor inclusive analysis
of data samples with enhanced heavy-flavor content.

The razor analysis is guided by studies of simulated events
generated with the \PYTHIA{6} \cite{Sjostrand:2006za} and
\MADGRAPH~\cite{madgraph} Monte Carlo programs, implemented using the
CMS \GEANTfour-based \cite{G4} detector simulation, and then
processed by the same software as that used to reconstruct
data.  Events with QCD multijets, top quarks, and electroweak bosons
are generated with \MADGRAPH interfaced with \PYTHIA for
parton showering, hadronization and the underlying event
description. To generate Monte Carlo samples for SUSY, the mass
spectrum is first calculated with \textsc{softsusy}~\cite{softsusy}
and the decays with \textsc{susyhit}~\cite{Susyhit}. The \PYTHIA{6}
program is used with the SLHA interface~\cite{SLHA} to generate
the events. 
Next-to-leading order (NLO) plus next-to-leading-logarithm (NLL)
cross section calculations are  used
~\cite{Beenakker:1996ch,Kulesza:2008jb,Kulesza:2009kq,Beenakker:2009ha,Beenakker:2011fu,NLO-NLL}.

\begin{figure}[htpb]
\begin{center}
\includegraphics[width=0.48\textwidth]{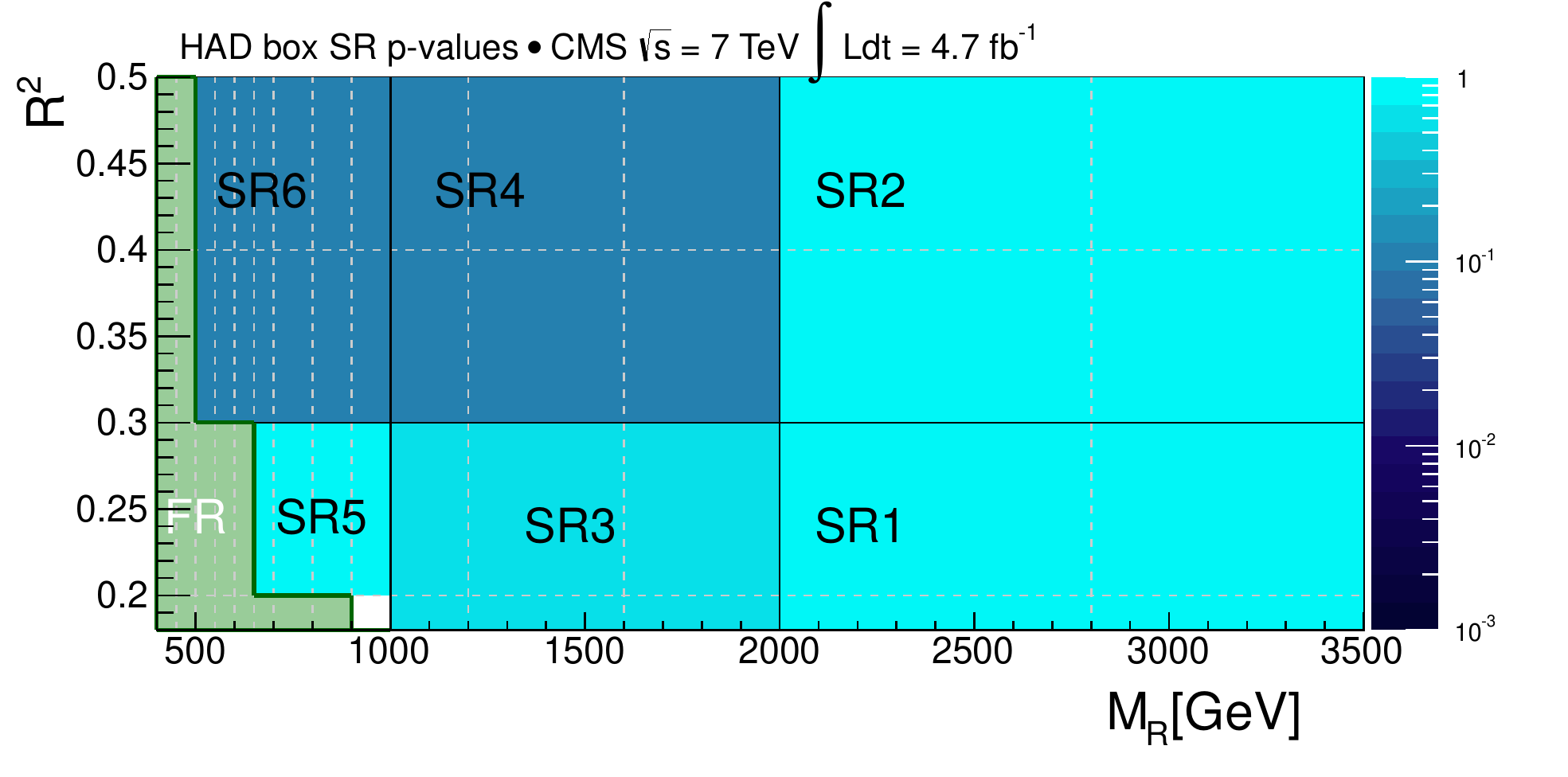}
\begin{scotch}{ l  c  c  c  c  c  }
\footnotesize
 \textbf{HAD}                  & 68\% range & mode & median & observed & p-value\\
 \hline
\textbf{SR1} & (0, 0.7) & 0.5 & 0.5 & 0 & 0.99 \\
\textbf{SR2} & (0, 0.7) & 0.5 & 0.5 & 0 &  0.99 \\
\textbf{SR3} & (45, 86) & 73 & 69 & 74 &  0.68\\
\textbf{SR4} & (4, 15) & 9.5 & 10.5 & 20 &  0.12 \\
\textbf{SR5} & (530, 649) & 566 & 593 & 581 &  0.82 \\
\textbf{SR6} & (886, 1142) & 987 & 1020 & 897 &  0.10 \\
\end{scotch}%
\caption{The p-values corresponding to the observed number of events
  in the HAD box signal regions (SR$i$). The green region indicates the fit
  region (FR) in the HAD box. Similar results are obtained for the
other boxes.
\label{fig:blue-plot}}
\end{center}
\end{figure}

For each of the main SM backgrounds, a control data sample is defined
from a subset of the data dominated by this particular background. It
is used to obtain a description of the shapes of the background
components from  data. In both simulation and control data
samples, the distributions of the major SM background events
(QCD multijets, $\ttbar+\text{jets}$, $\cPZ+\text{jets}$, and $\PW+\text{jets}$) are found to have a
simple exponential dependence on the razor variables $R^2$ and $M_\mathrm{R}$
over a large part of the $R^2$-$M_\mathrm{R}$ plane.  A full 2D SM background
representation is built using statistically independent data
samples. This representation used as input to the final fit performed in each
fit region (FR) defined by an L-shaped area in the $R^2$-$M_\mathrm{R}$
plane such as shown in Fig.~\ref{fig:blue-plot}.
The fit
region is dominated by SM processes. Signal contamination has only
a small impact on the determination of the background shape in the fit
region, as demonstrated through studies based on simulation.  The 2D
background model obtained in the fit region is extrapolated to the
rest of the $R^2$-$M_\mathrm{R}$ plane where the analysis is sensitive to a
potential SUSY signal.

The fit function has parameters describing the shapes and normalization of
the $R^{2}$ and $M_{R}$ distributions of the SM backgrounds. The
two-dimensional probability density function (pdf) $P_{j}(M_{R},R^{2})$ describing the $R^{2}$ versus
$M_{R}$ distribution of each considered SM process $j$ is found to be
well-approximated by instances of the same function $F_{j}(M_{R},R^{2})$:
\begin{equation}\begin{split}\label{eqn:2dpdf}
F_{j}(M_\mathrm{R}, R^2) =
\left[ k_{j}(M_\mathrm{R} - M^0_{R,j})(R^2 - R_{0,j}^2)-1\right] \times
 \\
\cmsSymbolFace{e}^{-k_{j}(M_\mathrm{R} - M^0_{R,j})(R^2 - R_{0,j}^2)}.
\end{split}\end{equation}
The scaling of the exponent as a function of the
thresholds on $M_\mathrm{R}$ and $R^2$ is described by the $k_j$ parameters
of the function. When integrated over $M_\mathrm{R}$ ($R^2$), this function recovers the
exponential dependence on $R^2$ ($M_\mathrm{R}$).

Each SM process in a given final-state box is
well described
by
a pdf $P_{j}$ that is the sum of a first and a second
component of the functional form  Eq.~\ref{eqn:2dpdf} with
separate normalizations. Studies of simulated events and fits to
control data samples with either a \cPqb-jet requirement or a \cPqb-jet veto
indicate that the parameters corresponding to the first
components of these backgrounds (with steeper slopes at low $M_\mathrm{R}$ and
$R^{2}$) are box-dependent. The parameters describing the second
component
are box-independent, and at the current precision of the
background model, they are identical among the dominant backgrounds
considered in these final states.

These sets of independent data control samples are used to derive
\textit{a priori} the background shape parameters.  The results are
incorporated in the final fits as a set of Gaussian penalty terms~\cite{penalty,Verkerke:2003ir} for the parameters $k_{j}$,
$M^0_{R,j}$ and $R_{0,j}^2$ multiplying the final likelihood
(Eq.(\ref{eq:Lb})). The RMS-values of the penalty terms for the $k_{j}$
parameters are typically $\sim$30\%.

An extended and unbinned maximum likelihood (ML) fit is performed in
each box using \textsc{RooFit}~\cite{Verkerke:2003ir}.  The fit 
performed in the fit region of the $R^2$-$M_\mathrm{R}$ plane 
provides the description of the SM background in the full
plane. The likelihood function for a given box is written as
\cite{Barlow:1990vc}:

\begin{equation}
\label{eq:Lb}
\mathcal{L}_{b} =  \frac{\cmsSymbolFace{e}^{-\left(\sum_{j \in SM} N_{j}\right)}}{N !}
\prod_{i=1}^{N} \left( \sum_{j \in SM} N_{j} P_{j}(M_{R,i},R^2_i)\right),
\end{equation}
where $N$ is the total event yield in the box, the sum runs over
all the SM  processes relevant for that box, and $N_{j}$ is
the yield of a given fit sample in the box.

The values of the shape parameters that
maximize the likelihood in these fits, along with the corresponding
covariance matrix, are used to define the background model and the
uncertainty associated with it. Additional background shape
uncertainties due to the choice of the functional form 
were found to be negligible \cite{razor-2012PRD}.

\begin{figure}[ht!]

\begin{center}
\includegraphics[width=0.45\textwidth]{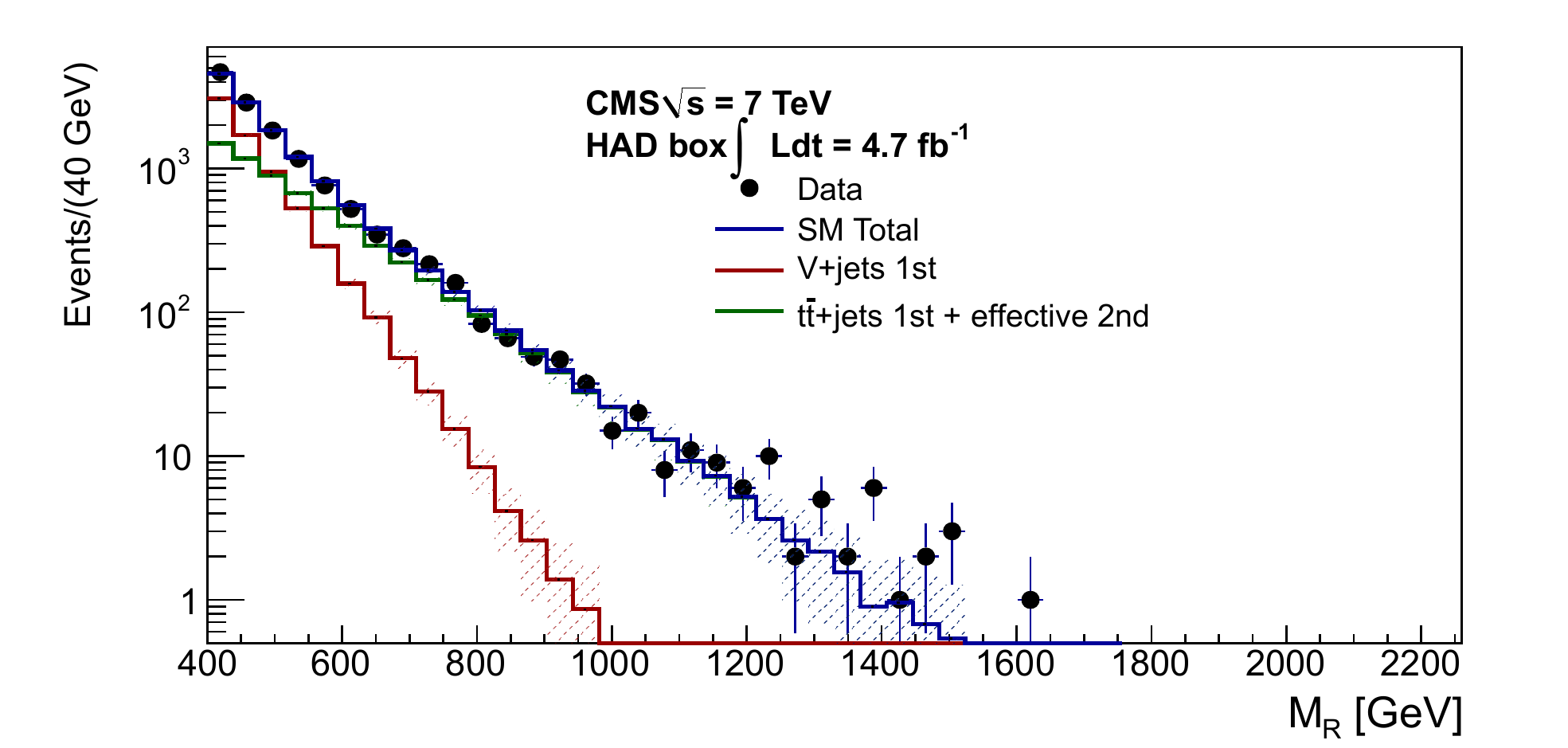}
\includegraphics[width=0.45\textwidth]{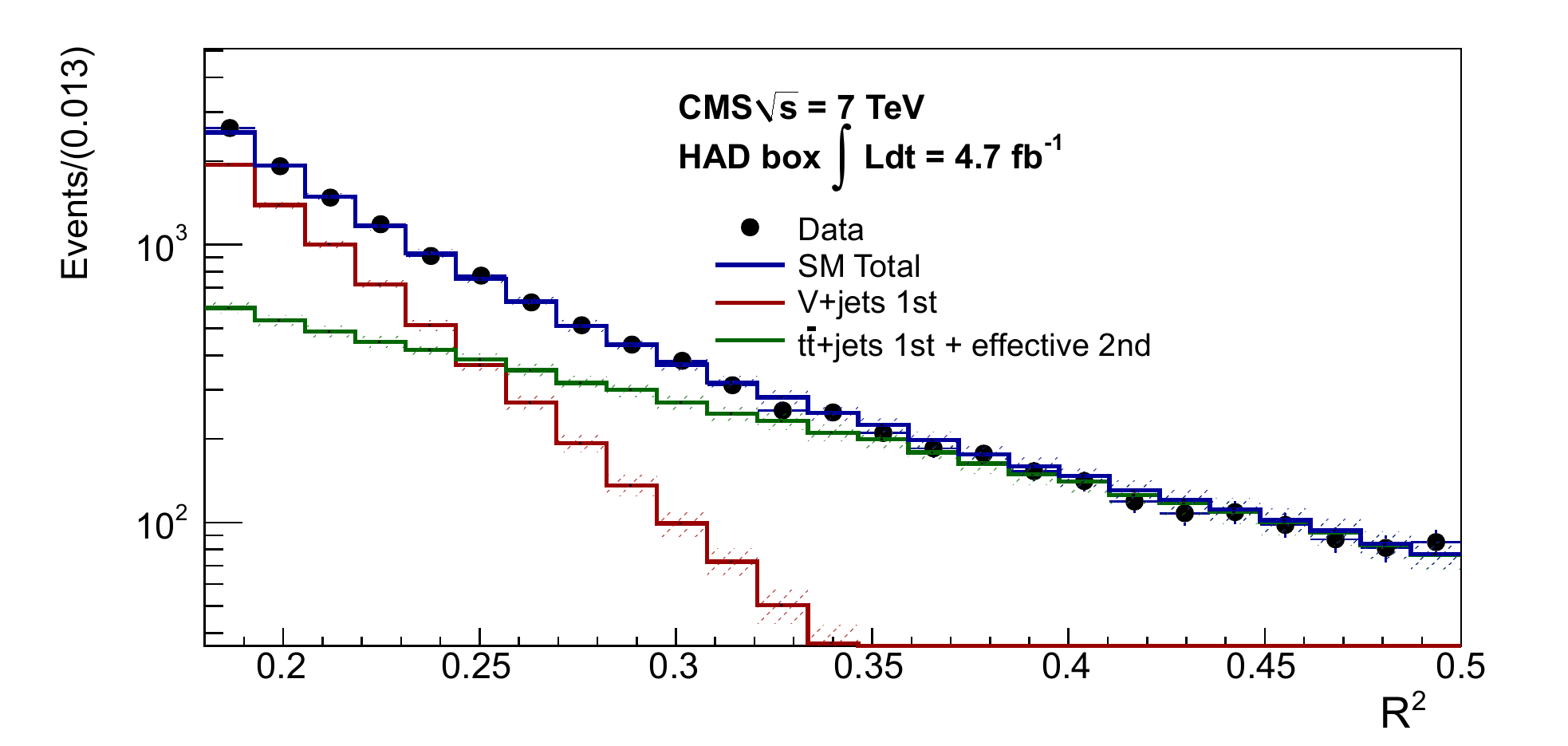}
\end{center}
\caption{Projection of the 2D fit result on $M_\mathrm{R}$ (\cmsLeft) and $R^2$
  (\cmsRight) for the HAD box. The blue histogram is the total standard
 model prediction as obtained from a single pseudo-experiment based on
 the 2D fit. The red and green histograms represent a steep-slope
 component denoted as $V+\text{jets}$ first component, and a component that
 encapsulates the steep-slope first component in $\ttbar + \text{jets}$ and
 the effective second component, which is indistinguishable for the
 different SM background processes. The fit is performed in the
 $R^2$-$M_\mathrm{R}$ fit region (FR as shown in Fig.~\ref{fig:blue-plot}) and projected into the full analysis
 region.  Only the statistical uncertainty band on the background
 predictions is drawn in these projections. \label{fig:had-box}}
\end{figure}

The result of the ML fit projected on
$M_\mathrm{R}$ and $R^2$ is shown in Fig.~\ref{fig:had-box} for
the HAD box.  No significant discrepancy is observed between the
data and the fit model in any of the six boxes \cite{razor-2012PRD}.

In order to establish the  compatibility of the background model with the observed dataset,
 we define six signal regions ($\mathrm{SR}{i}$) in the tail of the background distribution.
Using the background model returned by the ML fit, we derive the
distribution of the expected yield in each $\mathrm{SR}{i}$ using pseudo-experiments
accounting for correlations and uncertainties on the parameters
describing the background model. For each of the $\mathrm{SR}{i}$ the distribution
of the number of events derived by the pseudo-experiments is used to
calculate a two-sided p-value (as shown for the HAD box in
Fig.~\ref{fig:blue-plot}), corresponding to the probability of observing
an equal or less probable outcome for a counting experiment in each
signal region.   The p-values test the compatibility of the observed number
  of events in data with the SM expectation obtained from the
  background parameterization. We quote the median and the mode of the yield distribution for
each SR, together with the observed yield.

\begin{figure}[htpb]

\begin{center}
\includegraphics[width=0.45\textwidth]{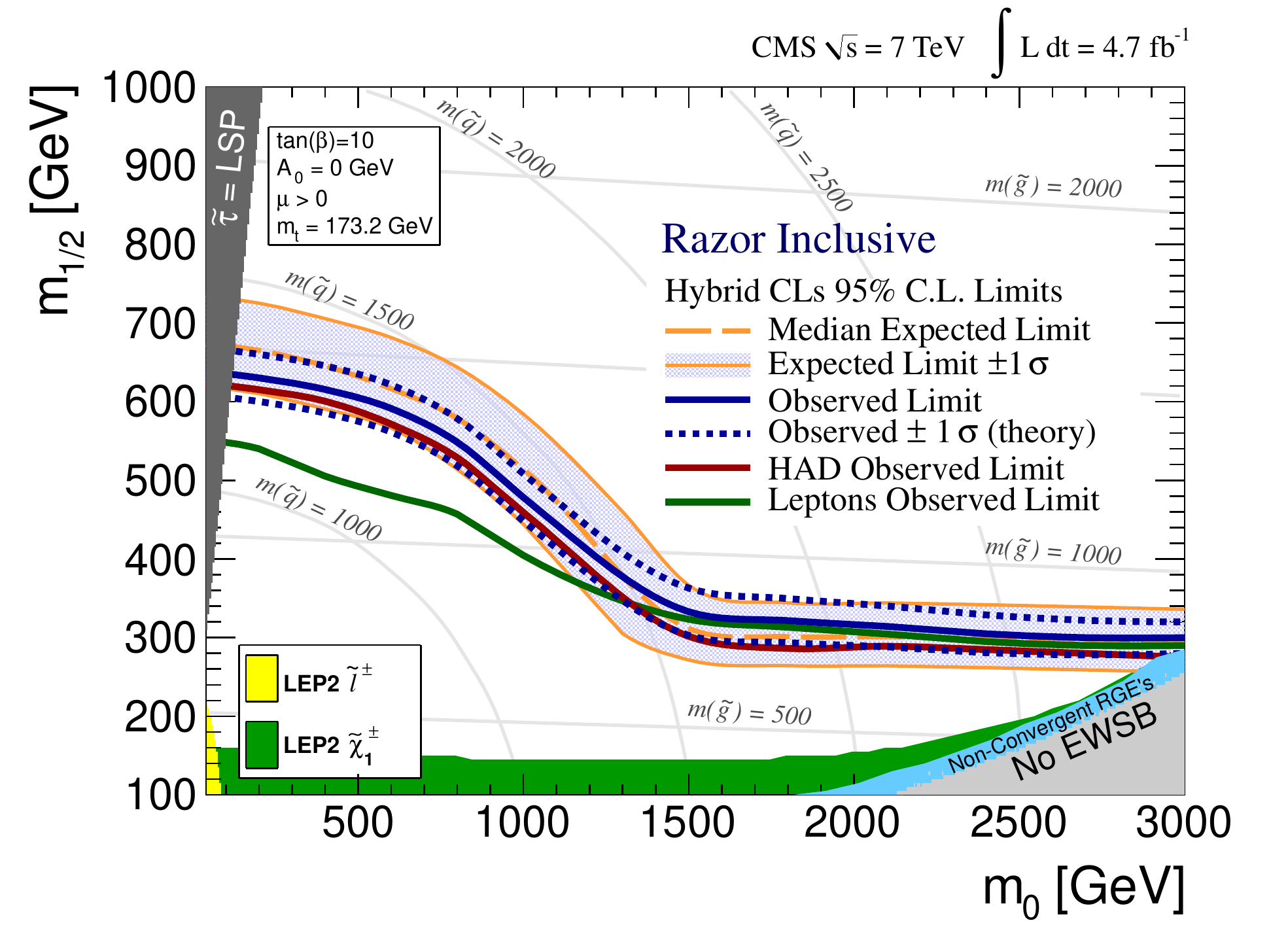}
\caption{Observed (solid blue curve) and median-expected (dashed
curve, shown with
  its $\pm$1 standard deviation uncertainty band) 95\% CL
limits in the ($m_{0}$, $m_{1/2}$) CMSSM plane (drawn according to~\cite{Rony}) with $\tan\beta=10$,
$A_{0} = 0$\GeV, and $\sgn(\mu) = +1$. Shown
separately are the observed HAD-only (solid crimson) and leptonic-only
(solid green) 95\% CL limits.
\label{fig:all-result}}
\end{center}
\end{figure}

For each box we consider the test statistic given by the logarithm of
the likelihood ratio $\ln{Q} = \ln\frac{\mathcal{
L}(s+b|H)}{\mathcal{L}(b|H)}$, where $H$ is the hypothesis under test;
$H_1$ (signal plus background) or the null hypothesis $H_0$
(background-only).
Given the distribution of $\ln{Q}$ for background-only and
signal-plus-background pseudo-experiments, and the value of $\ln{Q}$
observed in the data,
we calculate $\mathrm{CL}_\mathrm{s+b}$ and $1-\mathrm{CL}_\mathrm{b}$~\cite{Read:2002hq,Junk1999}. From these
values the $\mathrm{CL}_\mathrm{s}=\mathrm{CL}_\mathrm{s+b}/\mathrm{CL}_\mathrm{b}$ is computed for that model point. A
point in the constrained minimal supersymmetric standard model (CMSSM)
plane is excluded at 95\% confidence level (CL) if $\mathrm{CL}_\mathrm{s} < 0.05$. The
result is shown in Fig.~\ref{fig:all-result}.  The
shape of the observed exclusion curves reflect the changing relevant
SUSY strong production processes across the parameter space with
squark-squark and gluino-gluino production dominating at low and high
$m_0$, respectively.
The observed limit is less constraining than the median-expected limit
  at lower $m_0$ due to an excess of observed events in the HAD box at
  large $R^2$, where squark-pair production dominates over gluino-pair
  production.  Cascading decays of gluinos yield more leptons than
  decays of squarks.  Thus, relative to hadronic boxes, 
the contribution of lepton boxes
increases with $m_0$.

We estimate the systematic bias on the signal shape model due to
parton density functions  (point-by-point up to 30\%), jet energy scale (point-by-point up to 1\%) and lepton
identification (using using $\cPZ\to \ell^+\ell^-$ data, 1\% per lepton), and on the
signal yield  due to the luminosity uncertainty (2.2\%)~\cite{CMS-PAS-SMP-12-008},  the theoretical cross section
(point-by-point up to 15\%), razor trigger efficiency uncertainty (2\%), and  lepton trigger
efficiency uncertainty (3\%). In the \cPqb-tag analysis path an additional
systematic is considered for the
\cPqb-tagging efficiency (between 6\% and 20\% in $\pt$ bins~\cite{btag}).  We
consider variations of the function modeling the signal uncertainty
(log-normal versus Gaussian) as well as the $R^2$ and $M_\mathrm{R}$ binning choice, finding
negligible deviations in the result. 

\begin{figure}[htpb]
\begin{center}
\includegraphics[width=0.45\textwidth]{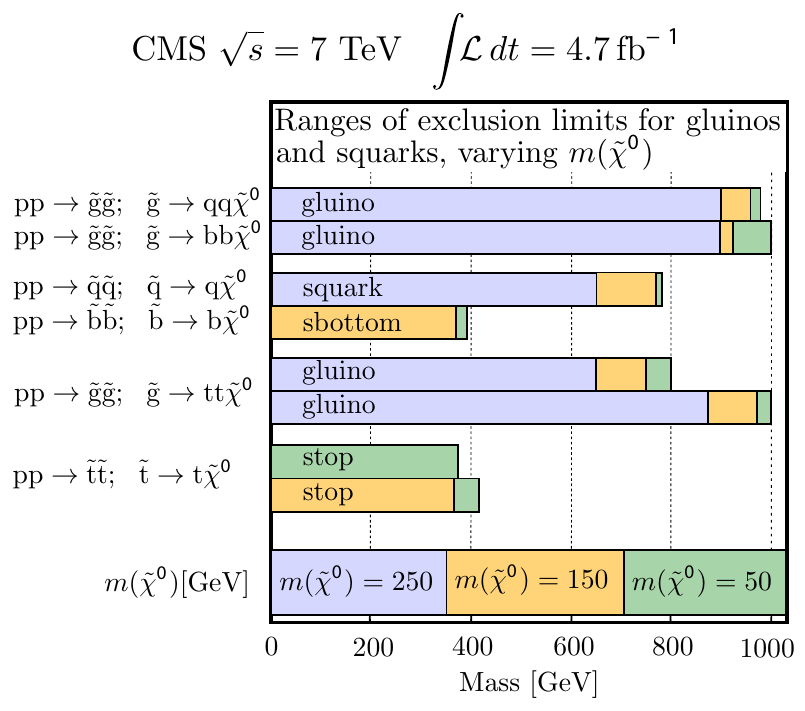}
\caption{Summary of the 95\% CL excluded largest parent mass as a function of the LSP mass in each
of the simplified models studied. Results from the inclusive
razor analysis (upper bars) and the \cPqb-jet razor analysis (lower
bars) are shown. \label{fig:sum-sms}}
\end{center}
\end{figure}

The results are also interpreted as cross section limits on a number
of simplified models~\cite{ArkaniHamed:2007fw}, where a limited set
of hypothetical particles and decay chains are introduced to produce a
given topological signature. Specific applications of these ideas have
appeared in Refs.~\cite{Alwall:2008ag,Alwall:2008va,Alves:2011wf}. For
each model studied, the excluded cross section at 95\% CL is derived as
a function of the mass of the produced particles (gluinos or squarks,
depending on the model) and the LSP mass, as well as the exclusion
curve corresponding to the NLL-NLO cross section.  Exclusion curves
for a factor of three cross section enhancement or reduction are also
produced as well as for $\pm$1 standard deviation variations in the NLL-NLO cross
section~\cite{razor-2012PRD}. Figure~\ref{fig:sum-sms} shows 
the 95\% CL~\cite{atlas-agreement} excluded largest parent mass as a function of
the LSP mass in each of the simplified model studies,
for both the inclusive and \cPqb-jet versions of the analysis.


In summary, we performed a search for squarks and gluinos using a data
sample of 4.7\fbinv of CMS data at $\sqrt{s} = 7\TeV$
proton-proton collisions in the razor variable space using a 2D shape
description of the relevant standard model processes.

No significant
excess over the background expectations is observed, and the results
are presented as a 95\% CL limit in the ($m_0$, $m_{1/2}$) CMSSM
parameter space. For
$m(\PSQ)\sim m(\PSg)$ we exclude squarks and gluinos up to
1.35\TeV,  and for $m(\PSQ)>m(\PSg)$ we
exclude gluinos up to 800\GeV. For simplified models we exclude up to
1\TeV for the gluino mass   and up to 800\GeV for the first and second generation
squark masses. For direct production of pairs of stops or sbottoms we
exclude stop and sbottom masses up to 400\GeV depending on the LSP
mass.

\section*{Acknowledgements}

We congratulate our colleagues in the CERN accelerator departments
for the excellent performance of the LHC and thank the technical and
administrative staffs at CERN and at other CMS institutes for their
contributions to the success of the CMS effort. In addition, we
gratefully acknowledge the computing centres and personnel of the
Worldwide LHC Computing Grid for delivering so effectively the
computing infrastructure essential to our analyses. Finally, we
acknowledge the enduring support for the construction and operation of
the LHC and the CMS detector provided by the following funding
agencies: BMWF and FWF (Austria); FNRS and FWO (Belgium); CNPq, CAPES,
FAPERJ, and FAPESP (Brazil); MEYS (Bulgaria); CERN; CAS, MoST, and
NSFC (China); COLCIENCIAS (Colombia); MSES (Croatia); RPF (Cyprus);
MoER, SF0690030s09 and ERDF (Estonia); Academy of Finland, MEC, and
HIP (Finland); CEA and CNRS/IN2P3 (France); BMBF, DFG, and HGF
(Germany); GSRT (Greece); OTKA and NKTH (Hungary); DAE and DST
(India); IPM (Iran); SFI (Ireland); INFN (Italy); NRF and WCU
(Republic of Korea); LAS (Lithuania); CINVESTAV, CONACYT, SEP, and
UASLP-FAI (Mexico); MSI (New Zealand); PAEC (Pakistan); MSHE and NSC
(Poland); FCT (Portugal); JINR (Armenia, Belarus, Georgia, Ukraine,
Uzbekistan); MON, RosAtom, RAS and RFBR (Russia); MSTD (Serbia); SEIDI
and CPAN (Spain); Swiss Funding Agencies (Switzerland); NSC (Taipei);
ThEPCenter, IPST and NSTDA (Thailand); TUBITAK and TAEK (Turkey); NASU
(Ukraine); STFC (United Kingdom); DOE and NSF (USA).

Individuals have received support from the Marie-Curie programme and
the European Research Council (European Union); the Leventis
Foundation; the A. P. Sloan Foundation; the Alexander von Humboldt
Foundation; the Belgian Federal Science Policy Office; the Fonds pour
la Formation \`a la Recherche dans l'Industrie et dans l'Agriculture
(FRIA-Belgium); the Agentschap voor Innovatie door Wetenschap en
Technologie (IWT-Belgium); the Ministry of Education, Youth and Sports
(MEYS) of Czech Republic; the Council of Science and Industrial
Research, India; the Compagnia di San Paolo (Torino); the Weston Havens
Foundation (US) and the HOMING PLUS programme of Foundation for Polish
Science, cofinanced from European Union, Regional Development Fund.

\bibliography{auto_generated}

\begin{thebibliography}{10}
\bibitem{atlas-agreement}
\hrefCMSnoop {} {}The summary results show the observed 95\% CL exclusions,
  based on the nominal NLO-NLL cross-sections. Since the interpolation is
  within the scan binning (in mass) we picked conservatively the limit close to
  the less 1$\sigma$ theoretical cross section in order to also facilitate
  comparisons an equal footing with ATLAS results.
\end{thebibliography}
\cleardoublepage \appendix\section{The CMS Collaboration \label{app:collab}}\begin{sloppypar}\hyphenpenalty=5000\widowpenalty=500\clubpenalty=5000\textbf{Yerevan Physics Institute,  Yerevan,  Armenia}\\*[0pt]
S.~Chatrchyan, V.~Khachatryan, A.M.~Sirunyan, A.~Tumasyan
\vskip\cmsinstskip
\textbf{Institut f\"{u}r Hochenergiephysik der OeAW,  Wien,  Austria}\\*[0pt]
W.~Adam, E.~Aguilo, T.~Bergauer, M.~Dragicevic, J.~Er\"{o}, C.~Fabjan\cmsAuthorMark{1}, M.~Friedl, R.~Fr\"{u}hwirth\cmsAuthorMark{1}, V.M.~Ghete, J.~Hammer, N.~H\"{o}rmann, J.~Hrubec, M.~Jeitler\cmsAuthorMark{1}, W.~Kiesenhofer, V.~Kn\"{u}nz, M.~Krammer\cmsAuthorMark{1}, I.~Kr\"{a}tschmer, D.~Liko, I.~Mikulec, M.~Pernicka$^{\textrm{\dag}}$, B.~Rahbaran, C.~Rohringer, H.~Rohringer, R.~Sch\"{o}fbeck, J.~Strauss, A.~Taurok, W.~Waltenberger, G.~Walzel, E.~Widl, C.-E.~Wulz\cmsAuthorMark{1}
% ...
\vskip\cmsinstskip
\dag:~Deceased\\
1:~~Also at Vienna University of Technology, Vienna, Austria\\
% ...

\end{sloppypar}
\end{document}